\documentclass[journal]{IEEEtran}
\usepackage{epsfig,euscript}
\usepackage[cmex10]{amsmath}

\begin{document}

\title{Phase-Aligned Space-Time Coding\\ for a Single Stream MIMO system}

\author{Joonsuk~Kim,~\IEEEmembership{Senior Member, ~IEEE}\\
\IEEEcompsocitemizethanks{\IEEEcompsocthanksitem Joonsuk Kim 
is with Office of the CTO,
Broadcom Corp., Sunnyvale, CA 94086,  E-mail: joonsuk@ieee.org| %,
}
}

\IEEEcompsoctitleabstractindextext{%
\begin{abstract}
We present a phase-aligned space-time coding scheme that expands the original Alamouti codeword to three or four transmit antennas ($N_t = 3$ or $4$)
with phase alignment. With $1 \sim 2$ bits feedback for the phase information, the fundamental performance penalty of
$10log_{10}(N_t)$ dB of orthogonal space-time coding compared to the optimum beamforming is reduced by 1 dB (for $N_t=3$) or 2 dB (for $N_t = 4$) on average.
With the proposed scheme, the full diversity order of $N_t$ is achievable, whereas the receiver architecture remains the same as the legacy
Alamouti decoding with codeword size of two, since the spatial expansion is transparent to the receiver. Our results show the proposed scheme outperforms
open-loop space-time coding for three or four transmit antennas by more than 3 dB.
\end{abstract}
}

\maketitle

\IEEEdisplaynotcompsoctitleabstractindextext
\IEEEpeerreviewmaketitle

\section{Introduction}
\indent
Space-Time Block Coding (STBC) is a keyword to obtain the diversity gain by using the multi-input multi-output (MIMO) system in digital communication. Since Alamouti codeword was proposed \cite{Alam9810} for two transmit antenna transmission, many transceiver techniques have been developed to realize the full diversity gain for more than two transmit antennas. However, the open-loop orthogonal transceiver 
design with full-rate and full diversity gain for arbitrary complex channels is proven to be limited to the case of two transmit 
antennas \cite{Taro9907}. For more than two antennas, many space-time coding schemes have been introduced to 
achieve the full diversity gain, however, with less than full-rate \cite{Xia0310}.

There was another attempt to achieve full-rate but with partial diversity by adopting quasi-orthogonal STBC (QO-STBC) structure \cite{Jafa0101}. While QO-STBC can extend Alamouti pairs to four transmit antennas easily, it often requires an extensive receiver
design, such as  Maximum-Likelihood (ML) technique \cite{Nagu9811_2}, to suppress the self-interference caused by non-orthogonal structure. However, with an aid of partial channel information feedback,
the full diversity gain is shown to be achievable for four transmit antenna system with rate one \cite{KimJ0711}. %KimJ0805
It also shows the feedback of only a few number of bits can achieve nearly full orthogonal STBC diversity gain, thereby cheaper 
minimum-mean-square-error (MMSE) receiver can match the performance of ML receiver.

However, there is still fundamental performance penalty, $10log_{10}(N_t)$ dB, even for orthogonal STBC \cite{KimJ0711, Mill0606}, compared to closed-loop beamforming, where $N_t$ is the number of transmit antennas.
In order to reduce the gap, it is recommended to obtain an aid of partial channel information combined with STBC with minimal feedback overhead, {\it e.g.,} $1 \sim 2$ bits, otherwise beamforming is still an attractive solution when larger feedback overhead is allowed \cite{Zhang1006}.
%It is because the transmitted symbols of STBC codeword are equally distributed over transmit antennas with unity power per symbol, %whereas the beamforming uses unity transmit power across all transmit antennas.
%For an example of $N_t = 4$, there is 6 dB penalty, even if the optimum orthogonal STBC scheme is used  \cite{KimJ0711, Mill0606}.
%There has been intensive research to reduce this gap with an aid of partial channel information available at the transmitting station %\cite{Liu0408,Jong0203}, but it requires as many number of bits as the beamforming feedback for comparable performance.
Transmit antenna selection can reach the same goal of full transmit diversity gain \cite{Toke0411,Kapi1006}, but it suffers from non-linearity distortion at high power transmission, which is a typical scenario to extend the transmission range with STBC.
Achieving full diversity gain by introducing new constellation for symbols \cite{Su0410} requires a special constellation optimized for re-designed STBC, which is not applicable for legacy systems with Gray code labeling \cite{Stuber96}.

In this paper, our goal is to improve an STBC scheme which can be detected by a legacy Alamouti decoder with codeword size of two, by introducing only a few number of bits for the feedback.
We achieve the goal with full transmit antenna diversity by
expanding STBC codewords over three or four transmit antennas. The spatial expansion matrix is found to align the phase of Alamouti pairs, whose information is obtained by channel
measurement and feedback performed by the receiver. We present a closed form solution for the phase alignment and 
show the proposed scheme
with  $1 \sim 2$ bits feedback performs to within only fractional dB of performance of the same system with full knowledge of the channel.
%As a result, the fundamental barrier of $10log_{10}(N_t)$ dB penalty compared to the beamforming is broken with marginal feedback %overhead. The constraint of non-time-varying channels over the size of codewords is retained as in the original Alamouti codeword
%whereas the diversity order is increased from two to four.
Note  \cite{KimJ0711} eliminates the off-diagonal interference terms in the effective channel, however the proposed scheme in this paper maximizes the diagonal terms in the effective channel since it retains orthogonality with the spatial expansion matrix.
The proposed scheme is transparent to the receiver, so a legacy Alamouti decoder  can still be applied without knowledge on whether this new scheme is applied at the transmitter.

\section{Phase-Aligned STBC}
\label{sec.PhA_STBC}
\begin{figure}[!t] %Hbt]
\centering
\includegraphics[height=0.9in,width=!]{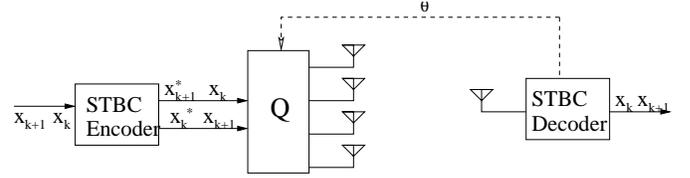}
\caption{The systeml diagram of phase-aligned STBC}
\label{fig.system}
\end{figure}

\indent
Alamouti codeword is a space-time coding scheme that can achieve full diversity gain when the number of transmit antennas is two.
When the number of transmit antennas is more than two, we can spread the output of STBC encoder over the $N_t$ space 
by adopting a spatial mapping matrix $Q$ with dimension of $N_t \times 2$ as shown in Figure \ref{fig.system}.
%with dimension of more than two. For an example with WLAN system, space-time streams after STBC encoder are expander over multiple %transmit antennas by a spatial mapping matrix, $Q$, to be transmitted over $N_t$ transmit antennas.
%Typically, $Q$ is an identity matrix for open-loop operation with STBC by randomly choosing an antenna pair, whereas $Q$ is a non-identity %unitary matrix for closed-loop beamforming operation with the data path bypassing the STBC encoder. % when the transmitter has full knowledge on the channel information. 
%In this paper, we consider an Alamouti pair encoded by STBC encoder by enabling a non-identity $Q$ matrix with limited channel information.
%
%

%With an STBC encoder, we consider an Alamouti codeword for a single stream: 
An Alamouti codeword for a single stream is
\begin{equation}
S = \left[
\begin{array}{cc}
x_{k} &  x_{k+1}^* \\
x_{k+1} & -x_{k}^* \\
\end{array}
\right], \label{eq.STBC_S}
\end{equation}
where row represents an antenna index and column represents a time instance for STBC. Superscript $^*$ denotes complex conjugate.
% and $^H$ denotes Hermitian transpose.
$x_{k}$ is the original input constellation symbol at $k^{th}$ original sequence.
%and $S$ is a $4 \times 2$ STBC/SFBC code matrix to be transmitted over 4 transmit antennas.
%Note two symbols in the code matrix $S$ in (\ref{eq.STBC_S})
%are multiplied by c which is equivalent to rotation by an angle $\theta$, where $c=e^{j\theta}$.
%In this paper, we assume that $\theta$ is a variable to be fed back in the general angle feedback scheme.

For more than two transmit antennas, we simply introduce a spatial mapping $Q$ after STBC encoder.
In order to preserve the orthogonality for the matched filter, {\it i.e.}, $Q^HQ=I$, where $I$ is an identity matrix,
it is easy to show that $(q)_{im}=0$ when $i+m$ is an odd integer without loss of generality, where $(q)_{im}$ is the $(i,m)^{th}$ component of $Q$.
With per-antenna power constraint to avoid non-linearity distortion at high transmit power, we can choose each element of $Q$ matrix with unity magnitude.
%following $Q$ matrix with a rotation factor, $e^{-j\theta}$:

We consider four transmit antennas as a simple example. With a $4 \times 2$ spatial mapping matrix $Q$, we can distribute the Alamouti
codeword over 4 transmit antennas. The spatial mapping matrix $Q$ we choose is
\begin{equation}
Q = \left[
\begin{array}{cccc}
1 & 0 & e^{-j\theta} & 0 \\
0 &  1 & 0 &  e^{-j\theta}
\end{array}
\right]^T,
\label{eq.Q}
\end{equation}
where $\theta$ is determined by the channel information and superscript $^T$ denotes transpose. Herein, the scale factor is moved to the channel in (\ref{eq.Hm}) below.
Equivalently, the new STBC codeword with $Q$ matrix to be transmitted over four transmit antennas is
%satisfying (\ref{eq.y=Hx+n}) - (\ref{eq.Q}) to be transmitted over the air is
\begin{equation}
S' = \left[
\begin{array}{cccc}
x_{k} & x_{k+1} & e^{-j\theta}x_{k} &  e^{-j\theta}x_{k+1} \\
x_{k+1}^* & -x_{k}^* &  e^{-j\theta}x_{k+1}^* & -e^{-j\theta} x_{k}^* \\
\end{array}
\right]^T.
 \label{eq.STBC_S4}
\end{equation}

 Assuming the channel is not varying over the size of the codeword, {\it i.e.}, $h_{im}(k) = h_{im}(k+1)$  where $k$ is a time index, the received signal for the STBC codeword in (\ref{eq.STBC_S}) with $N_r$ receive antennas is given by

\begin{equation}
\begin{aligned}
\left[
\begin{array}{c}
R_1\\
R_2\\
\vdots\\
R_{N_r}
\end{array}
\right] &= 
\left[
\begin{array}{c}
H_1 \\
H_2\\
\vdots \\
H_{N_r}
\end{array}
\right]
\times 
\left[
\begin{array}{c}
x_{k} \\
x_{k+1}
\end{array}
\right]
+ \left[
\begin{array}{c}
N_1 \\
N_2 \\
\vdots \\
N_{N_r}
\end{array}
\right] \\
&= {\mathcal H}  \underbar{x} + \underbar{n},
\end{aligned} \label{eq.y=Hx+n}
\end{equation}
%\fontsize{10}\selectfont
where 
\begin{equation}
\begin{aligned}
\mathcal{H} & = \left[ \begin{array}{cccc}H_1^T & H_2^T & \cdots &  H_{Nr}^T \end{array}\right]^T \\
H_m & = \frac{1}{\sqrt{N_t}} \left[\begin{array}{cc} h_{1m} + e^{-j\theta}h_{3m}& h_{2m}  + e^{-j\theta}h_{4m} \\ 
-h_{2m}^* - e^{j\theta}h_{4m}^* & h_{1m}^* +e^{j\theta} h_{3m}^* \end{array} \right] \\
R_m &= \left[\begin{array}{cc} r_{m,k} & r_{m,k+1}^* \end{array} \right]^T \\
N_m &= \left[\begin{array}{cc} n_{m,k} & n_{m,k+1}^* \end{array}\right]^T.
\end{aligned}
\label{eq.Hm}
\end{equation} 
$r_{m,k}$ and $n_{m,k}$ are the received signal and noise at $m^{th}$ receive antenna at the $k^{th}$ time instance, and $N_t = 4$.
%$h_{ij}$ is the channel from the $i^{th}$ transmit antenna to the $j^{th}$ receive antenna for the $1^{st}$ stream, $x_{1,k}$, and $g_{ij}$ is the channel from the $(i+2)^{th}$ transmit antenna to the $j^{th}$ receive antenna for the $2^{nd}$ stream, $x_{2,k}$. 
$h_{im}$ is the channel from the $i^{th}$ transmit antenna to the $m^{th}$ receive antenna, without the time index $k$.
%Both  $h_{ij}$ and  $g_{ij}$ are scaled by $1/\sqrt{N_t}$ for unity transmit power with 4 transmit antennas ($N_t = 4$).
$R_m$ and $N_m$ are a $2 \times 1 $ received signal vector and a $2 \times 1 $ noise vector, respectively, at the $m^{th}$ receive antenna.
%${\bf H}$ is a $2 N_r \times 4$ effective channel matrix with a group of 2 adjacent time instance or frequency tones.

%Then, each pair of rows in the effective channel, ${\mathcal H}_m = H_m \times Q$, from the transmitter to the $m^{th}$ receive 
%antenna is
%\begin{equation}
%{\mathcal H}_m = \left[
%\begin{array}{cc}
%h_{1m}+e^{-j\theta}h_{3m} & e^{-j\theta}h_{2m} + h_{4m} \\
%-h_{2m}^* -e^{-j\theta}h_{4m}^*& e^{-j\theta}h_{1m}^* +h_{3m}^*
%\end{array}
%\right].
%\end{equation}

Note $H_m$ preserves the orthogonality since the off-diagonal terms of $H_m^H H_m$ are all zeros, where $^H$ denotes Hermitian transpose. Then the codeword can be optimally detected by left-multiplying the receive signal vector in (\ref{eq.y=Hx+n}) by ${\mathcal H}^H$; 

\begin{equation}
{\mathcal H}^H {\mathcal H} = \displaystyle\sum_{m=1}^{N_r} H_m^HH_m = \frac{1}{N_t}
\left[
\begin{array}{cc}
\Sigma  & 0 \\
0 & \Sigma
\end{array}
\right], \label{eq.HH_Nr}
\end{equation}
where 
\begin{equation}
\begin{aligned}
\Sigma &= \displaystyle\sum_{m=1}^{N_r}\left( | h_{1m}+e^{-j\theta}h_{3m}  |^2 + |h_{2m}+e^{-j\theta}h_{4m} |^2\right) \\
&= \displaystyle\sum_{i=1}^{4} \sum_{m=1}^{N_r} |h_{im}|^2   + 2\Re\{e^{-j\theta}\alpha\}    %{\text{,   }}
\end{aligned}
\label{eq.Sigma}
\end{equation}
and
\begin{equation}
\alpha = \displaystyle\sum_{m=1}^{N_r}\left(h_{1m}^*h_{3m} + h_{2m}^*h_{4m}\right).
\label{eq.alpha}
\end{equation}
Herein, $\Re\{\cdot\}$ denotes the real part of the argument.

For other antenna configuration, we can use the similar concept. For an example of $N_t=3$, (\ref{eq.Q}) becomes
\begin{equation}
Q = \left[
\begin{array}{ccc}
1 & 0 & e^{-j\theta} \\
0 &  1 & 0 
\end{array}
\right]^T,
\label{eq.Q2}
\end{equation}
thereby the last row of $S'$ in (\ref{eq.STBC_S4}) does not exist.
Also, (\ref{eq.Hm}) and (\ref{eq.Sigma}) can be rewritten as
\begin{equation}
H_m = \frac{1}{\sqrt{N_t}}\left[\begin{array}{cc} h_{1m} + e^{-j\theta}h_{3m}& h_{2m}   \\ 
-h_{2m}^*  & h_{1m}^* +e^{j\theta} h_{3m}^* \end{array} \right] 
\label{eq.Hm2}
\end{equation}
and
\begin{equation}
\begin{aligned}
\Sigma &= \displaystyle\sum_{m=1}^{N_r}\left( | h_{1m}+e^{-j\theta}h_{3m}  |^2 + |h_{2m} |^2\right) \\
&= \displaystyle\sum_{i=1}^{3} \sum_{m=1}^{N_r} |h_{im}|^2   + 2\Re\{e^{-j\theta}\alpha\},    %{\text{,   }}
\end{aligned}
\label{eq.Sigma2}
\end{equation}
where $\alpha$ in (\ref{eq.Sigma2}) is
\begin{equation}
\alpha = \displaystyle\sum_{m=1}^{N_r}\left(h_{1m}^*h_{3m} \right).
\label{eq.alpha2}
\end{equation}

Note that the effective channel has a nice orthogonal structure; there is no need to operate additional interference cancellation process. 
%Then the rotation $\theta$
%is found to maximize the magnitude of $\Sigma$. Therefore, it is simply found to match the phase to $\alpha$ as
Finally, the closed form solution for $\theta$ to maximize the magnitude of $\Sigma$ by matching the phase of $\alpha$ is
\begin{equation}
\theta = \angle{\alpha},
\label{eq.theta}
\end{equation}
where $\angle\{\cdot\}$ denotes the phase of the argument.

In practice with a digital communication system, the phase information
$\theta$ to be fed back needs to be quantized. The feedback information can be obtained by quantizing the angle within $[0, 2\pi ]$ range,
and the most simplest form with 1-bit or 2-bits feedback can be found as $e^{-j\theta} = 1$ or $-1$, or $e^{-j\theta} = 1, -1, j$ or $-j$,
respectively, which are found by quantizing the angle in  (\ref{eq.theta}).

With this proposed scheme that maintains the orthogonal coding structure, we can simply apply a matched filter approach to detect the signal. However, to combat with noise-enhancement at low SNR, where most of STBC schemes are compelling to extend the range of transmission with high diversity order,
the receiver filter is typically designed with an MMSE solution with noise power of $\sigma^2$ \cite{Stuber96}.

When $N_t>4$, as long as $N_t$ is an even number, we can easily expand this scheme with another angles, $\theta_n$ with $n=2,3,\cdots,N_t/2-1$, by adding multiple pairs of Alamouti codeword over the transmit antennas. When $N_t$ is an odd number, the same expansion of even number of transmit antennas with $N_t+1$ is applied, but  the last channel response, $h_{N_t+1,m}$ is replaced with zero. To generalize this scheme for arbitrary configuration is out of scope for this paper.
%Although it may not be possible to find a closed form solution for $\theta_n$, we can find the quantized $\theta_n$ by brute-force search %with criterion to co-phase with channel products as in (\ref{eq.Sigma}) and  (\ref{eq.alpha}).

\section{Simulation and Comparison} 
\label{sec.sims}
\indent
%As seen in (\ref{eq.HH}) and (\ref{eq.HH_Nr}), STBC/SFBC does not obtain full orthogonality which results in interference between symbols within STBC/SFBC codewords. In order to suppress such interference, it is required for the linear receiver to employ zero-forcing (ZF) or minimum mean-square error (MMSE). The system performance gain with angle feedback can be easily estimated by measuring the output signal-to-noise ratio (SNR) after interference cancellation. 

\begin{figure}[!t] %Hbt]
%\leavevmode
%\centerline{\psfig{figure=SNR_output_4x2_SFBC.eps,height=2.8in,width=!}}
\centering
\includegraphics[height=2.8in,width=!]{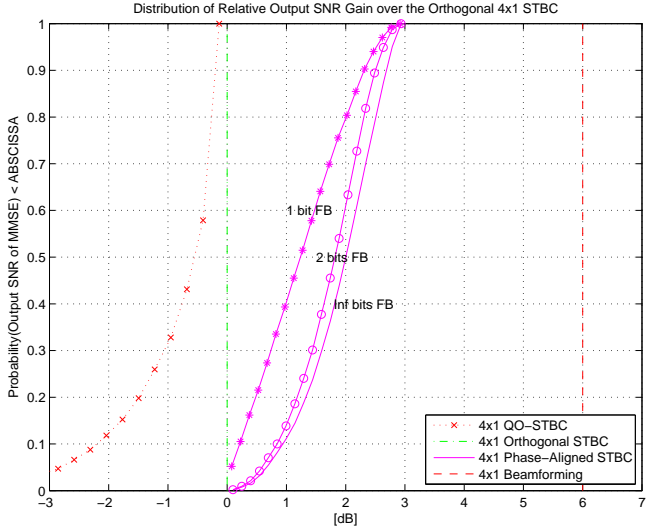}
\caption{Distribution of MMSE output SNR gain with respect to the orthogonal STBC receiver}
\label{fig.outputSNR_HH}
\end{figure}

We consider a 4-transmit, 1-receive antenna ($4 \times 1$) system with independent and identically distributed (i.i.d.) complex Gaussian channels to demonstrate the benefits of the proposed scheme. It is well known that beamforming has $10\log_{10}(N_t)$ dB gain over an orthogonal STBC scheme, since symbols at each transmit antenna has unity power for STBC while the sum of signal power for all transmit antennas is unity for beamforming. For a $4 \times 1$ system, singular value decomposition (SVD) is used for beamforming and \cite{KimJ0711} is used for orthogonal STBC scheme as the optimum solution, assuming full channel information is available at the transmitter.

Figure \ref{fig.outputSNR_HH} shows the output  signal-to-noise ratio (SNR) gain of an MMSE receiver over the orthogonal STBC scheme  \cite{KimJ0711} when the noise power is 10 dB less than the received signal power. 
It demonstrates exactly 6 dB difference between the SVD beamforming and the orthogonal STBC scheme.
%, assuming full channel information is available at the transmitter. 
With QO-STBC \cite{Jafa0101}, 
%where non-zero off-diagonal terms exists in the effective channel, 
it shows performance penalty compared to orthogonal STBC due to the non-orthogonal structure with interference. 
On the other hand, the proposed phase-aligned STBC shows the output SNR gain from 0 dB to 3 dB over orthogonal STBC;
%. When the channels are randomized, we don't expect any gain if ($h_1$, $h_3$) and ($h_2$, $h_4$) pairs are orthogonal each other %($\alpha = 0$), but obtain the maximized gain up to 3 dB if ($h_1$, $h_3$) and ($h_2$, $h_4$) pairs happen to be co-phased (each element of $\alpha$ is real), from (\ref{eq.Sigma}).
the maximum gain is obtained when two channel products are naturally co-phased to begin with, or no gain when even and odd
channel pairs are out of phase ($\alpha = 0$).
Interestingly, with 2-bits feedback on quantized value of $\theta$, the output SNR curve is very tight to the curve with full knowledge of the channel. Since beamforming typically requires more than 18 bits feedback to nearly achieve the performance with infinite bits feedback \cite{KimJ0605,IEEE80211n}, the phase-aligned STBC with 2-bits feedback has only 4 dB penalty on average for the output SNR with 90$\%$ percentile saving on the feedback overhead, compared to beamforming.

\begin{figure}[!t] %Hbt]
%\leavevmode
%\centerline{\psfig{figure=SNR_output_4x2_SFBC.eps,height=2.8in,width=!}}
\centering
\includegraphics[height=2.8in,width=!]{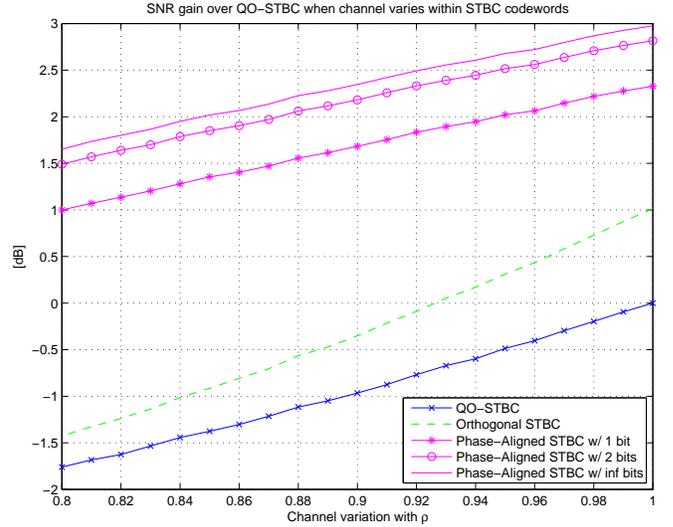}
\caption{Sensitivity of MMSE output SNR gain with respect to non-fading QO-STBC, when channels vary within the STBC codewords}
\label{fig.outputSNR_Sense}
\end{figure}

Most of research on the topic of STBC assumed the channel does not change within the STBC codeword. While it is a reasonable assumption for slow fading channels, it becomes a more strong assumption when the size of the STBC codewords is larger.
 Note QO-STBC  \cite{Jafa0101} and orthogonal STBC  \cite{KimJ0711} assume $h_{im}(k+l)= h_{im}(k)$ for $l=1,2$ and 3, but the proposed scheme assumes it only for $l=0$ and 1, because it maintains the size of codewords to two. In reality, however, the channel response can vary within a codeword, which results in self-interference in the matched filter operation.
Figure \ref{fig.outputSNR_Sense} demonstrates how much sensitive the MMSE performance of the proposed scheme is, compared with other STBC schemes,
when the channel is varying within the STBC codeword. 
In this plot, the correlation $\rho$ represents the difference of the channel at each time instance; 
$h_{im}(k+1) = \rho h_{im}(k) + \sqrt{1-\rho^2} g_{im}(k)$ where both of $h_{im}(k)$ and $g_{im}(k)$ are i.i.d. Gaussian random variables.
Then, the MMSE output SNR is calculated and the gap to the output SNR of QO-STBC with non-fading channels is plotted.
%The plots show the SNR gain over QO-STBC when $\rho=1$, which indicates the channel does not vary within the codeword.
When $\rho$ decreases from 1, all of STBC schemes suffer from self-interference. Especially, the impact on orthogonal STBC is
larger due to additional loss on the stale feedback information. However, the proposed scheme maintains the loss less than 1.5 dB when $\rho=0.8$,
which is smaller than QO-STBC and orthogonal STBC with codeword size of four.

\begin{figure}[!t]%Hbt]
\centering
%\includegraphics[height=2.8in,width=!]{PER_.eps}
%\caption{PER for $4 \times 2$ STBC with angle feedback for a 11n channel D, Non-Line-of-Sight (NLOS), with QPSK and convolutional coding r=3/4}
\includegraphics[height=2.8in,width=!]{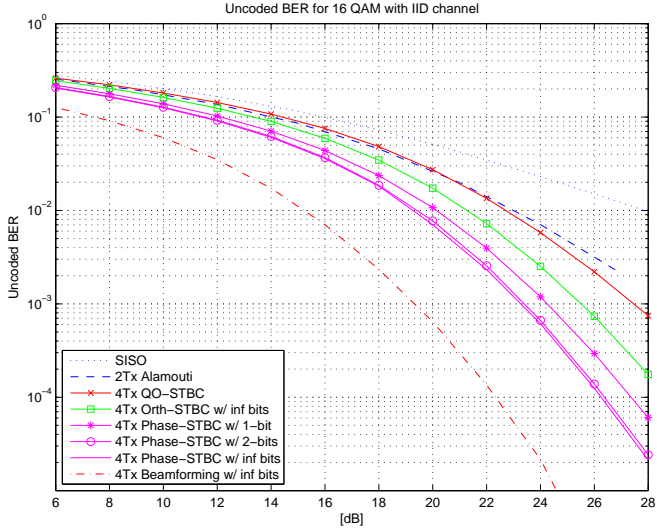}
\caption{Uncoded BER performance comparison for i.i.d. complex Gaussian channel with 16 QAM for a $4 \times 1$ system}
\label{fig.uBER_4x1_16Q_IID}
\end{figure}

Figure \ref{fig.uBER_4x1_16Q_IID} shows the average bit-error-rate (BER) performance comparison when the information bits are not coded. For this simulation, 16 quadrature amplitude modulation (QAM) is used with i.i.d. complex Gaussian channels. It demonstrates  beamforming, the proposed scheme and orthogonal STBC achieves full diversity order of four transmit antennas. 
%, but QO-STBC does not. 
The proposed scheme reduces the gap to beamforming by 2 dB, compared with orthogonal STBC.
% has 2 dB gain over the orthogonal STBC, thereby reducing the fundamental gap of 6 dB to the beamforming.
$1 \sim 2$ bits quantization performs to within less than 1 dB of the optimum performance with full resolution feedback.

\begin{figure}[!t]%Hbt]
\centering
%\includegraphics[height=2.8in,width=!]{PER_.eps}
%\caption{PER for $4 \times 2$ STBC with angle feedback for a 11n channel D, Non-Line-of-Sight (NLOS), with QPSK and convolutional coding r=3/4}
\includegraphics[height=2.8in,width=!]{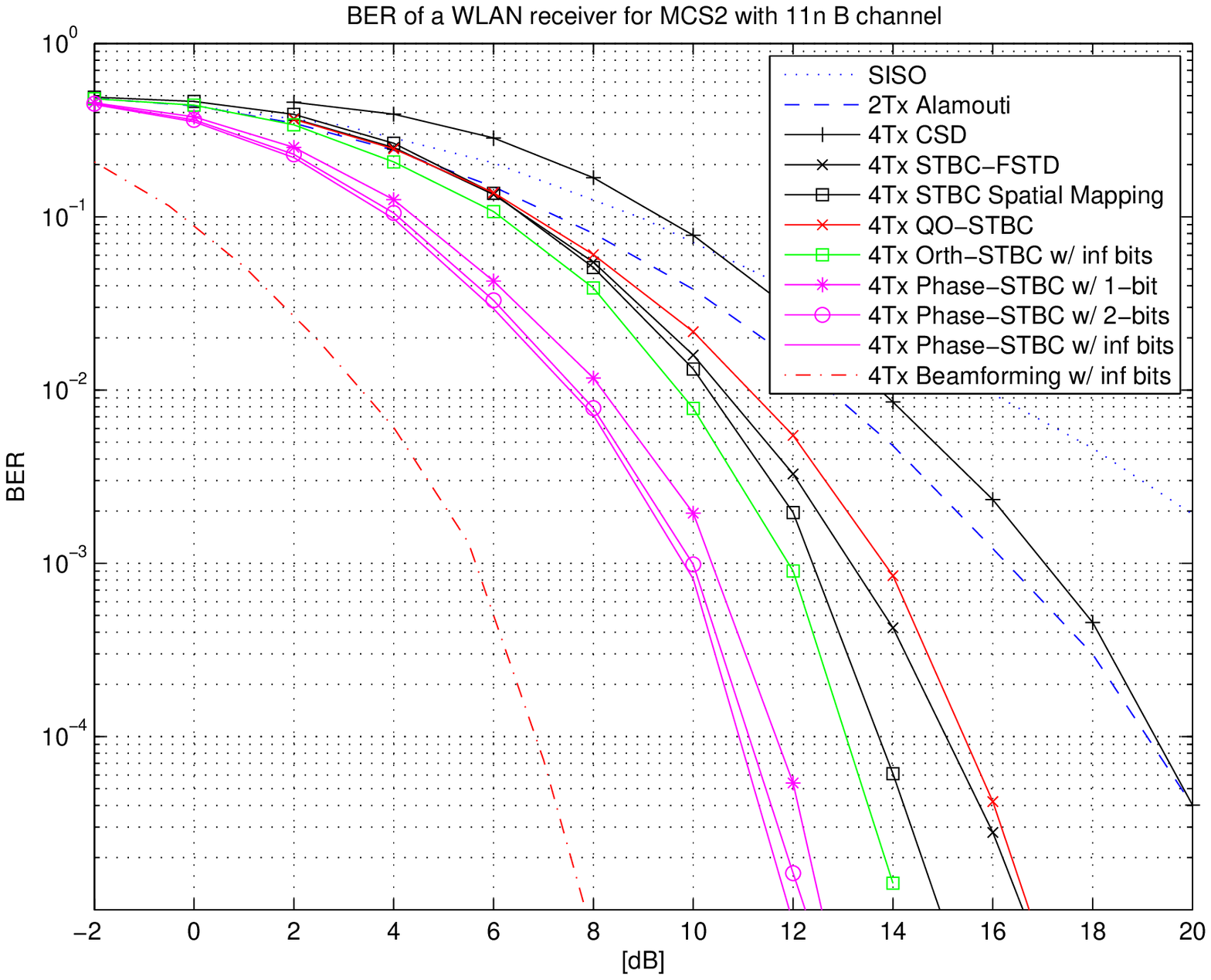}
\caption{BER performance of Phase-Aligned STBC for 802.11n channel model B with 4 QAM and convolutional coding r=3/4 for a $4 \times 1$ system}
\label{fig.BER_2Qr34_11nB}
\end{figure}

We also simulate the average BER performance with a coding for a $4 \times 1$ 
system based on the orthogonal frequency-division multiplexing (OFDM) link model for 802.11n system \cite{IEEE80211n}.
%In the simulation, each packet is 1000 information bytes long for 20 MHz bandwidth with 52 data tones where 64 FFT size is chosen.
The MMSE receiver is employed with an assumption of perfect channel estimation and convolutional codes. 
Figure \ref{fig.BER_2Qr34_11nB} shows the BER results as a function of the average received SNR 
for 802.11n channel B which has 25 $n$sec rms delay spread with 0.5 antenna correlation \cite{IEEE80211n_chan}.
For reference, the BER performance of single-input single output (SISO), $2 \times 1$ Alamouti STBC, $4 \times 1$ with 
cyclic-shift delay (CSD) are also plotted, where all transmit symbols with CSD and STBC schemes are scaled properly with $1/\sqrt{N_t}$
for fair comparison. In the 802.11n system \cite{IEEE80211n}, CSD values are fixed to $[0, -50, -100, -150] nsec$ for four transmit antennas.
In addition,  Frequency Switched Transmit Diversity (FSTD) \cite{LTE10} is simulated. STBC-FSTD basically chooses two transmit antennas out of available transmit antennas for transmission with an alternating fixed fashion over tones; the first and the second transmit antennas for tone 1, the first and the third transmit antennas for tone 2, and so on. This scheme typically looks for frequency selecetive effects to obtain more coding gain and saves transmit power by half by nulling the other antennas. STBC with spatial mapping is the proposed scheme without angle information feedback; $\theta$ in (\ref{eq.Q}) is fixed to zero.
Finally, the proposed phase-aligned STBC is simulated with 1-bit, 2-bits and infinite bits (full resolution) angle feedback
to take into account the quantization effect, while both beamforming based on SVD and orthogonal-STBC are assumed to have
full channel knowledge at the transmitter for the best performance.

\begin{figure}[!t]%Hbt]
\centering
%\includegraphics[height=2.8in,width=!]{PER_.eps}
%\caption{PER for $4 \times 2$ STBC with angle feedback for a 11n channel D, Non-Line-of-Sight (NLOS), with QPSK and convolutional coding r=3/4}
\includegraphics[height=2.8in,width=!]{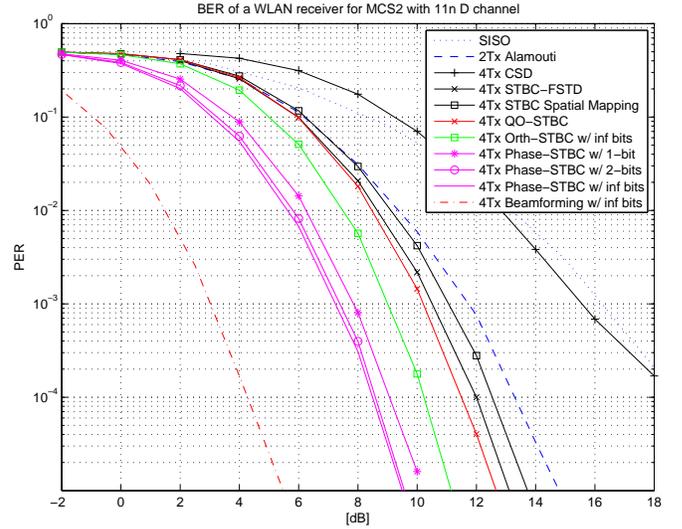}
\caption{BER performance of Phase-Aligned STBC for 802.11n channel model D with 4 QAM and convolutional coding r=3/4 for a $4 \times 1$ system}
\label{fig.BER_2Qr34_11nD}
\end{figure}

The simulation results show that the proposed scheme with 2-bits feedback nearly achieves the BER performance with infinite bits angle feedback, which has 2 dB gain over orthogonal-STBC and 4 dB loss to beamforming on average. 
%Compared to SVD beamforming, which requires a lot of feedback bits \cite{love0410}, infinite bits angle feedback STBC suffers from the power penalty of 3 dB\footnote{Unity power for transmitted signals are spread over streams for SVD beamforming, {\it i.e.}, $x$ needs to be scaled by $1/\sqrt{2}$, while the power needs to be spread over the transmit antennas for STBC, {\it i.e.}, $x$ is scaled by $1/\sqrt{4}$.} with additional 0.2 dB %0.7 dB 
%loss due to residual interference in off-diagonal terms of $H^HH$ in (\ref{eq.HH_Nr}).
Compared to existing open-loop solutions, {\it e.g.,} CSD and QO-STBC schemes, it has more than 4 dB gain.

Figure  \ref{fig.BER_2Qr34_11nD} shows BER performance results for more frequency selective channels, 802.11n channel D, which has 50 $n$sec rms delay spread with 0.3 antenna correlation \cite{IEEE80211n_chan}.
It also demonstrates the proposed scheme has 4 dB loss to beamforming and 2 to 3 dB gain over orthogonal STBC and QO-STBC.
Note the performance of CSD scheme is worse in this plot, compared to results with results in Figure \ref{fig.BER_2Qr34_11nB}, since the CSD scheme loses its gain with frequency selective channels.

\begin{figure}[!t]%Hbt]
\centering
%\includegraphics[height=2.8in,width=!]{PER_.eps}
%\caption{PER for $4 \times 2$ STBC with angle feedback for a 11n channel D, Non-Line-of-Sight (NLOS), with QPSK and convolutional coding r=3/4}
%\includegraphics[height=2.8in,width=!]{uncodedBER_16QAM.eps}
\includegraphics[height=2.8in,width=!]{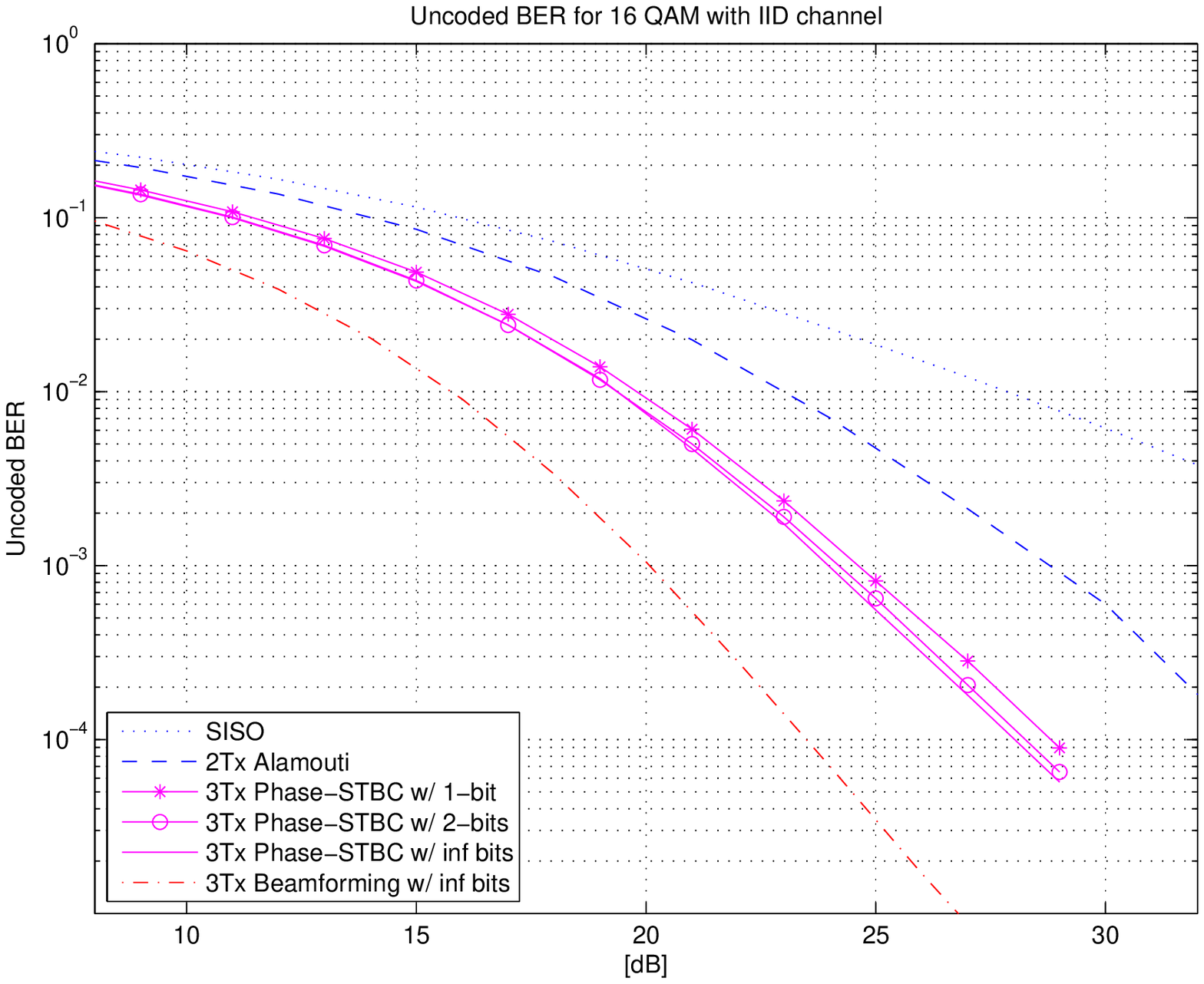}
\caption{Uncoded BER performance comparison for i.i.d. complex Gaussian channel with 16 QAM for a $3\times1$ system}
\label{fig.uBER_3x1_16Q_IID}
\end{figure}

\begin{figure}[!t]%Hbt]
\centering
%\includegraphics[height=2.8in,width=!]{PER_.eps}
%\caption{PER for $4 \times 2$ STBC with angle feedback for a 11n channel D, Non-Line-of-Sight (NLOS), with QPSK and convolutional coding r=3/4}
\includegraphics[height=2.8in,width=!]{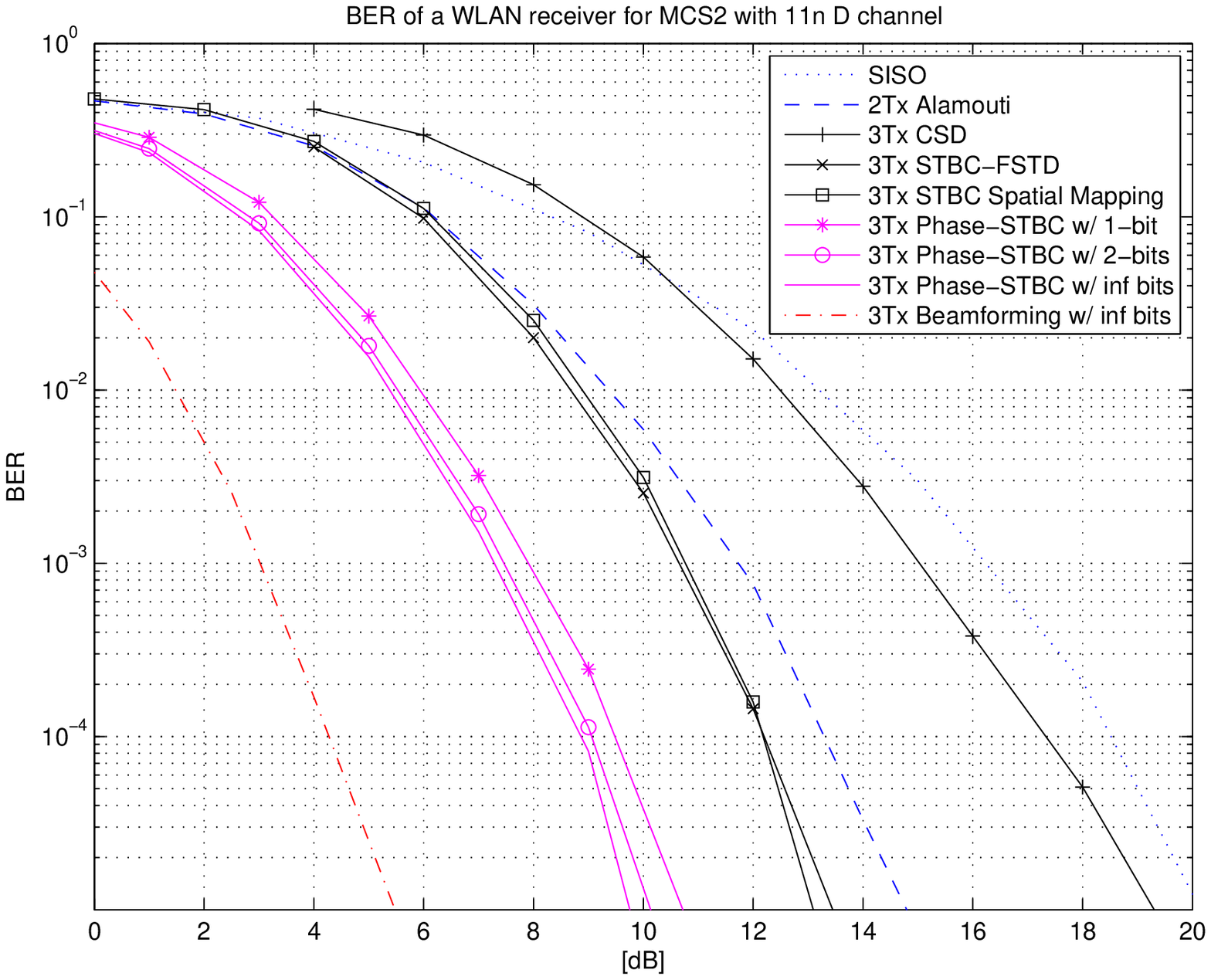}
\caption{BER performance of Phase-Aligned STBC for 802.11n channel model D with 4 QAM and convolutional coding r=3/4 for a $3 \times 1$ system}
\label{fig.BER_3x1_2Q34_11nD}
\end{figure}

In order to demonstate the performance of the proposed scheme for other antenna configuration, performance simulation for a $3 \times 1$ system is also performed. Figure \ref{fig.uBER_3x1_16Q_IID} shows the BER performance of an uncoded system with 16 QAM and i.i.d. complex Gaussian channels. We observe significant performance improvement by employing the proposed phase-aligned STBC scheme. For a realistic system, Figure \ref{fig.BER_3x1_2Q34_11nD} shows the BER performance of 4 QAM and convolutional coding rate of 3/4 with 802.11n channel D. With three transmit antennas, CSD values are fixed to $[0, -100, -200] nsec$, but orthogonal STBC and QO-STBC are not available. The results show the fundamental gap of $10log_{10}(3)$ dB to the ideal beamforming is reduced by 1 dB, and the proposed scheme has the performance gain of 3 dB over open-loop schemes.

\section{Conclusions}
\label{sec.conc}
\indent
We presented phase-aligned space-time coding scheme that spreads Alamouti codewords over the space with phase rotation.
The rotation factor is found to compensate the phase of sum of channel products to maximize the diversity gain.
The proposed phase-aligned scheme outperforms orthogonal STBC  \cite{KimJ0711}, QO-STBC  \cite{Jafa0101} and STBC-FSTD \cite{LTE10} by 2 to 4 dB for a $4 \times 1$ antenna configuration. For a $3 \times 1$ system, it outperforms STBC-FSTD by 3 dB.
The proposed scheme with 3 or 4 transmit antennas also maintains the performance loss of 4 dB to beamforming with 90\% feedback overhead saving.
%and reduce the gap to the beamforming by 2 dB.
Simulation for an i.i.d. complex Gaussian channel (flat fading) and 802.11 channel B and D (frequency selective fading)
shows that the 2-bits feedback scheme  performs to within a few fractional dB of %0.2 dB 
the performance with infinite bits feedback.
%, therefore saving $90\%$ percentile of feedback at the cost of 4 dB performance penalty compared to the beamforming.
Note the proposed scheme is applied only to the transmitting station, but does not require any new detection algorithm on the legacy Alamouti decoder with codeword size of two, except partial channel feedback capability.

%\include{PhA_STBC_4x1_arxivs}
%\bibliographystyle{IEEEtran}

\end{document}